\documentclass[prd]{revtex4}

\usepackage{bm}
\usepackage{epsfig}
\usepackage{amssymb}
\usepackage{amsmath}
\usepackage{slashbox}
\usepackage{calrsfs}
\usepackage{multirow}
\usepackage{rotating}

\begin{document}

\def\be{\begin{equation}}
\def\ee{\end{equation}}

\title{Analytic Continuation of Harmonic Sums}
\author{S.\ Albino}
\affiliation{{II.} Institut f\"ur Theoretische Physik, Universit\"at Hamburg,\\
             Luruper Chaussee 149, 22761 Hamburg, Germany}

\begin{abstract}
We present a method for calculating any (nested) harmonic sum 
to arbitrary accuracy for all complex values of the argument.
The method utilizes the relation between harmonic sums and (derivatives of)
Hurwitz zeta functions,
which allows a harmonic sum to be calculated as an expansion valid for large values of its argument.
A program for implementing this method is also provided.
\end{abstract}


\maketitle

\section{Introduction}

Probably the most efficient method for evaluating cross sections involving parton distribution functions
and/or fragmentation functions involves the inverse Mellin transform of the Mellin transformed cross section,
because the DGLAP equation can be solved analytically in Mellin space up to the desired order in 
the strong coupling $\alpha_s$ \cite{Furmanski:1981cw}.
In a minimal subtraction scheme, this evolution is determined by perturbatively calculable, mass independent, 
timelike and/or spacelike splitting functions respectively,
which depend on (nested) harmonic sums \cite{Blumlein:1998if} in Mellin space \cite{Moch:1999eb,Moch:2004pa},
as do massless coefficient functions \cite{Moch:1999eb} and other quantities in perturbation theory.
Harmonic sums are defined for positive integer values of the Mellin space variable $N$ as
\be
\begin{split}
S_{k_1,k_2,k_3,\ldots}(N)
=\sum_{n_1=1}^N \frac{({\rm sgn}(k_1))^{n_1}}{n_1^{|k_1|}}\sum_{n_2=1}^{n_1}\frac{({\rm sgn}(k_2))^{n_2}}{n_2^{|k_2|}}
\sum_{n_3=1}^{n_2}\frac{({\rm sgn}(k_3))^{n_3}}{n_3^{|k_3|}}\ldots
\end{split}
\label{defofSk1k2forintN_prev}
\ee
However, in the inverse Mellin transform, $N$ is complex.
The numerical evaluation of harmonic sums in this case is the purpose of this letter,
and is intended for important calculations such as the cross section calculations
of global fits of parton distribution functions (see e.g.\ Ref.\ \cite{Nadolsky:2008zw}) 
and fragmentation functions (see e.g.\ Ref.\ \cite{deFlorian:2007aj}).

The complex analysis of harmonic sums has been understood for some time \cite{Blumlein:1997vf}.
Algebraic relations among harmonic sums, namely
their quasi-shuffle algebra \cite{MEHoffman}, worked out to weight 6 in Ref.\ \cite{Blumlein:2003gb},
and structural relations \cite{Blumlein:2009fz},
reduces the number of harmonic sums that needs to be calculated,
and can be used to relate them through polynomials to $S_1(N)$,
which fixes their single poles at non-positive integer values of N \cite{Blumlein:2007dj}.
The analytic continuation is then unique \cite{CarlsonTitchmarsh}.
The continued form of harmonic sums that the method of this letter relies on is the large $|N|$, or asymptotic, expansion
\be
S_{k_1,k_2,k_3,\ldots}(N)=\sum_{k=0}^\infty \sum_{l=0}^\infty A_{k_1,k_2,k_3,\ldots}^{(l),k}
\left(\ln N +\gamma_E\right)^l  \frac{1}{N^k}.
\label{SeriesforSk2k3n}
\ee
(We choose to expand in $\left(\ln N +\gamma_E\right)$ instead of $\ln N$ since this simplifies 
explicit expressions somewhat.)
The coefficients $A_{k_1,k_2,k_3,\ldots}^{(l),k}$ consist of rational numbers and special 
numbers such as $\ln(2)$, $\zeta_n$, $\pi^n$ etc.
They may also depend on $\eta$, where $\eta=1$ if the harmonic sum is continued from even integer values of $N$
and $\eta=-1$ if continued from odd values
(see Ref.\ \cite{Kazakov:1987jk,Kotikov:2005gr} for detailed studies of $\eta$).
To achieve the expansion in eq.\ (\ref{SeriesforSk2k3n}), we extend the method introduced in the appendix of Ref.\ \cite{Albino:2005me}
for the calculation of $\tilde{S}=S_{-2,1}$. 
There this harmonic sum was analytically continued by extending 
the sum over $n_1$ in eq.\ (\ref{defofSk1k2forintN_prev}) to infinity.
This approach was also used in Ref.\ \cite{Kotikov:2005gr}.
In Ref.\ \cite{Albino:2005me}, the harmonic sum was then calculated as an expansion in inverse powers of $N$.
There are 2 problems with this approach that prevent it from being applied to other harmonic sums.
Firstly, an intermediate step involved the identification of a non convergent series with a finite result,
which cannot work for all harmonic sums.
Secondly, the analytic continuation only works when $k_1 \neq 1$, since otherwise the harmonic sum is 
logarithmically singular as $N\rightarrow \infty$.
These two problems are solved in the present letter, respectively
by identifying certain sums with the
Hurwitz zeta functions, whose expansion for large complex arguments is well known,
and reordering the sum when $k_1=1$ such that this identification also works in this case.
We note that a method for obtaining the large $|N|$ expansion for harmonic sums was also given in 
Ref.\ \cite{CoestermansEnjalbertMinhPetitot} using the isomorphism of harmonic sums to shuffle algebra.

Our approach provides an alternative to that of Ref.\ \cite{Blumlein:1997vf,Blumlein:2005jg},
where approximations were given for the harmonic sums involved in various perturbative quantities such as the 
3-loop anomalous dimension. In these references, the harmonic sums
are determined semi-analytically by calculating the analytic Mellin transforms of the 
relevant $x$-space functions which are partly calculated by polynomial approximations
using the very accurate minimax method, which leads to an accuracy of $10^{-10}$ to $10^{-12}$.
(In fact, only some harmonic sums were determined this way, since the remaining can then be determined by using
the various relations between harmonic sums discussed above, some of which 
are presented in Ref.\ \cite{Blumlein:1998if}).
Here we are concerned with deriving a
method for the numerical evaluation of any harmonic sum anywhere in complex Mellin space to any desired accuracy.

The rest of the letter is organized as follows. 
In section \ref{ancon}, the basic analytic continuation that allows 
harmonic sums to be calculated for complex arguments is discussed.
The improvement to this result that leads to fast, accurate calculations via an expansion in 
the inverse of the complex argument is discussed in section \ref{largeNk1ne1}.
Section \ref{largeNk1eq1} discusses the additional manipulations required when 
the first index of the harmonic sum is equal to 1, but the method itself is essentially the same.
The additional calculation required when the absolute value of the argument is not sufficiently large 
for the expansion to work is given in section \ref{Nnotlarge}.
Finally, in section \ref{summary} we summarize our approach in full.
The appendix contains explicit results for harmonic sums whose weight 
(the sum of the absolute values of the indices) is less than or equal to 3.

\section{Analytic continuation \label{ancon}}

Our approach is most conveniently derived by writing eq.\ (\ref{defofSk1k2forintN_prev}) in the form
\be
S_{k_1,k_2,k_3,\ldots}(N)
=\sum_{n_1=1}^N \frac{({\rm sgn}(k_1))^{n_1}}{n_1^{|k_1|}} S_{k_2,k_3,\ldots}(n_1).
\label{defofSk1k2forintN}
\ee
Equation (\ref{defofSk1k2forintN}) can be analytically continued to complex values of $N$ as follows: 
The sum over $n_1=1,2,\ldots,N$ is replaced with 
a sum over $n_1=1,2,\ldots,\infty$, and then the additional contribution due to this replacement is canceled by 
subtracting a second term which is identical except with the sum over $n_1=N+1,N+2,\ldots,\infty$.
Then, in this second term, $N$ is subtracted from $n_1$ so that the sum is over $n_1=1,2,\ldots,\infty$.
The result is
\be
S_{k_1,k_2,k_3,\ldots}(N)=S_{k_1,k_2,k_3,\ldots}(\infty)
-\sum_{n_1=1}^\infty \frac{({\rm sgn}(k_1))^{n_1 +N}}{(n_1+N)^{|k_1|}} S_{k_2,k_3,\ldots}(n_1+N).
\label{defofSk1k2forcompN}
\ee
Consequently, for any complex $N$, 
all harmonic sums can be determined from the $S_k$, being the harmonic sums whose depth (the number of indices) is equal to 1.

Although eq.\ (\ref{defofSk1k2forcompN}), which was also derived in Ref. \cite{Albino:2005me,Kotikov:2005gr},
is essentially our starting point, as it stands it is insufficient 
for the numerical evaluation of harmonic sums because when $k_1 \neq 1$ the sum over $n_1$ may converge slowly,
as mentioned in Ref.\ \cite{Blumlein:2005jg},
while for $k_1=1$ the series for $N=\infty$ in eq.\ (\ref{defofSk1k2forintN}) does not converge.
As we will see, the improvement in this letter is the development of a method for obtaining the large $|N|$ 
expansion of any harmonic sum, which converges fast for sufficiently large $|N|$.

\section{Calculation for large $|N|$ when $k_1 \neq 1$ \label{largeNk1ne1}}

In this section, we give the general procedure for obtaining the large $|N|$ expansion 
of any harmonic sum, eq.\ (\ref{SeriesforSk2k3n}),
from the expansion of harmonic sums of lower depth and weight.
This means that all harmonic sums' expansions can be obtained once the expansion for depth 1 harmonic sums are known.
In fact, it suffices to know only the expansion for $S_1(N)$ \cite{Blumlein:1998if},
\be
\begin{split}
S_1(N)=&\psi(N+1)+\gamma_E,\\
\end{split}
\label{Snintermsofpsiderivs}
\ee
where $\psi(N)$ and its derivatives can be evaluated for large $|N|$ using
\be
\psi(N+1)=\ln(N)+\frac{1}{2N}-\sum_{k=1}^\infty \frac{B_{2k}}{2k N^{2k}},
\label{psiatlargeN}
\ee
since the expansion for the remaining harmonic sums of depth 1, which are also presented in Ref.\ \cite{Blumlein:1998if}, 
can be determined from eq.\ (\ref{Snintermsofpsiderivs}) using eq.\ (\ref{defofSk1k2forcompN}):
We find
\be
S_k(N)=-\frac{d}{dN}S_{k-1}(N)+(k-1)\zeta(k)\ {\rm for}\ k\geq 2,
\ee
so that
\be
S_k(N)=\frac{(-1)^{k-1}}{(k-1)!} \psi^{(k-1)} (N+1)+\zeta(k)\ {\rm for}\ k\geq 2.
\label{Skcont}
\ee
Using the result
\be
S_{-k}(N)=\sum_{n=1}^\infty \frac{(-1)^n}{n^k}
+\frac{(-1)^N}{2^k}\left[S_k\left(\frac{N}{2}\right)-S_k \left(\frac{N-1}{2}\right)\right]\ {\rm for}\ k\geq 1
\ee
that follows from eq.\ (\ref{defofSk1k2forintN_prev}), one obtains from eq.\ (\ref{Snintermsofpsiderivs}) 
the $S_{-k}$ for $k\geq 1$:
\be
\begin{split}
S_{-1}(N)=&(-1)^N \beta(N+1)-\ln 2,\\
S_{-k}(N)=&\frac{(-1)^{k-1}}{(k-1)!}(-1)^N \beta^{(k-1)}(N+1)-\left[1-\frac{1}{2^{k-1}}\zeta(k)\right]\ {\rm for}\ k\geq 2
\label{Snintermsofpsiderivs2}
\end{split}
\ee
where $\beta(N)=(1/2)\left[\psi((N+1)/2)-\psi(N/2)\right]$.

We now show how eq.\ (\ref{defofSk1k2forcompN}) can be expanded in $1/N$
once the expansion in eq.\ (\ref{SeriesforSk2k3n}) for $S_{k_2,k_3,\ldots}(N)$ is known.
We assume $k_1 \neq 1$, and will deal with the case $k_1=1$ in section \ref{largeNk1eq1}.
The expansion for $S_{k_1,k_2,k_3,\ldots}(N)$ may then be calculated by inserting
eq.\ (\ref{SeriesforSk2k3n}) with $N\rightarrow n_1 +N$  into eq.\ (\ref{defofSk1k2forcompN})
and making the identifications
\be
\begin{split}
\sum_{n_1=1}^\infty \frac{1}{(n_1+N)^x}&=\zeta(x,N+1)\\
\sum_{n_1=1}^\infty \frac{(-1)^{n_1}}{(n_1+N)^x}&=\frac{1}{2^x}\left[\zeta\left(x,\frac{N+1}{2}+\frac{1}{2}\right)
-\zeta\left(x,\frac{N+1}{2}\right)\right]\\
\sum_{n_1=1}^\infty \ln^m(n_1+N)\frac{(\pm 1)^{n_1}}{(n_1+N)^x}
&=\left(-\frac{d}{dx}\right)^m\sum_{n_1=1}^\infty \frac{(\pm 1)^{n_1}}{(n_1+N)^x},
\label{genzetaident}
\end{split}
\ee
where the Hurwitz zeta function $\zeta(x,N+1)$ is set equal to its large $|N|$ expansion
\be
\zeta(x,N+1)=N^{-x} \left[\frac{N}{x-1}-\frac{1}{2}
+\sum_{k=1}^\infty \frac{B_{2k}}{(2k)!} \frac{(2k+x-2)!}{(x-1)!}\frac{1}{N^{2k-1}}\right],
\label{expanzeta}
\ee
with $B_{2k}$ the Bernoulli numbers.
Note that the second result in eq.\ (\ref{genzetaident}) follows from the first.

Equation (\ref{expanzeta}) can be derived from the result
\be
\zeta(x,N+1)=\frac{\psi^{(x-1)}(N+1)}{(-1)^x (x-1)!},
\ee
which follows from the first result in eq.\ (\ref{genzetaident})
together with eq.\ (\ref{Skcont}) and the analytic continuation in eq.\ (\ref{defofSk1k2forcompN}) applied 
to $S_k$ for $k\geq 2$ (see also Ref.\ \cite{Abramowitz:1968}),
and the result that follows from eq.\ (\ref{psiatlargeN}):
\be
\psi^{(x-1)}(N+1)=(-1)^x \left[\frac{(x-2)!}{N^{x-1}}-\frac{(x-1)!}{2N^x}+\sum_{k=1}^\infty B_{2k}
\frac{(2k+x-2)!}{(2k)! N^{2k+x-1}}\right].
\ee
Equation (\ref{expanzeta}) is continued to all real values of $x>1$, noting that
$\zeta(x,N+1)$ must be real in this case for real values of $N>-1$.

In many cases, the constant $S_{k_1,k_2,k_3,\ldots}(\infty)$ in eq.\ (\ref{defofSk1k2forcompN}) 
can be expressed in terms of natural numbers
by identifying the corresponding series in eq.\ (\ref{defofSk1k2forintN_prev}) when $N=\infty$.
Where this is not possible, it may be obtained numerically by comparing the expansion of eq.\ (\ref{defofSk1k2forcompN}) in $1/N$,
once that has been determined by the method of this section,
with the result obtained from eq.\ (\ref{defofSk1k2forintN_prev}), for large, integer $N$.

\section{Calculation for large $|N|$ when $k_1 = 1$ \label{largeNk1eq1}}

Some modification to the way the harmonic sum is calculated, eq.\ (\ref{defofSk1k2forintN_prev}), is required
before the above approach, namely the expansion of eq.\ (\ref{defofSk1k2forcompN}) at large $|N|$, 
can be applied when $k_1=1$, because in this case $S_{1,k_2,k_3,\ldots}(N)$ grows as (powers of) $\ln N$ as $N$ increases.
If $k_2 \neq 1$, the replacement
\be
\sum_{n_1=1}^N\sum_{n_2=1}^{n_1}f(n_1,n_2)=\sum_{n_1=1}^N \sum_{n_2=1}^N f(n_1,n_2)+\sum_{n_1=1}^N f(n_1,n_1)
-\sum_{n_1=1}^N\sum_{n_2=1}^{n_1}f(n_2,n_1)
\label{replacesum}
\ee
in the definition in eq.\ (\ref{defofSk1k2forintN})
gives an expression from which a large $|N|$ expansion can be obtained:
\be
S_{1,k_2,k_3,\ldots}(N)=S_1 (N) S_{k_2,k_3,\ldots}(N)+S_{{\rm sgn}(k_2) (|k_2|+1),k_3,\ldots}(N)
-\sum_{n_1=1}^N \frac{({\rm sgn}(k_2))^{n_1}S_1 (n_1)}{n_1^{|k_2|}}S_{k_3,\ldots}(n_1).
\label{s1k2k3}
\ee
The expansion of the last term in eq.\ (\ref{s1k2k3}) is achieved
by performing the usual analytic continuation of section \ref{ancon} and expansion of section \ref{largeNk1ne1}.
However, if $k_2=1$, then the replacement of sums in eq.\ (\ref{replacesum}) can be used again
in the last term in eq.\ (\ref{s1k2k3}).
Repeating this process $p$ times for a harmonic sum in which the first $p$
indices equal 1 and $k_{p+1}\neq 1$ gives 
\be
\begin{split}
S_{\underbrace{\scriptstyle 1,\ldots,1}_p ,k_{p+1},\ldots}(N)=&\sum_{n=1}^{p-1} (-1)^{n-1}
S_{\underbrace{\scriptstyle 1,\ldots,1}_n}(N) S_{\underbrace{\scriptstyle 1,\ldots,1}_{p-n} ,k_{p+1},\ldots}(N)\\
&+\sum_{n=1}^{p-1} F^{n-1,2}_{\underbrace{\scriptstyle 1,\ldots,1}_{p-n-1},k_{p+1},\ldots}(N)
+F^{p-1,1}_{k_{p+1},\ldots}(N).
\end{split}
\label{masterdecompofsp1}
\ee
where we define (we define the harmonic sum with no indices to be equal to one, i.e.\ $S=1$)
\be
F^{n,r}_{k_1,k_2,\ldots}(N)=(-1)^n \sum_{n_1=1}^N \frac{({\rm sgn}(r))^{n_1}}{n_1^{|r|}}
S_{\underbrace{\scriptstyle 1,\ldots,1}_n}(n_1)S_{k_1,k_2,\ldots}(n_1).
\label{defFnrks}
\ee
$F^{n,r}_{k_1,k_2,\ldots}(N)$ can be analytically continued and expanded at large $|N|$
according to the method of section \ref{largeNk1ne1} only when $r\neq 1$, which excludes the last term
in eq.\ (\ref{masterdecompofsp1}).
Fortunately, we can derive from it the results we need to do this.
Firstly, by performing the replacement of sums in eq.\ (\ref{replacesum}) on the last term,
eq.\ (\ref{masterdecompofsp1}) becomes
\be
\begin{split}
S_{\underbrace{\scriptstyle 1,\ldots,1}_p ,k_{p+1},\ldots}(N)=&\sum_{n=1}^p (-1)^{n-1}
S_{\underbrace{\scriptstyle 1,\ldots,1}_n}(N) S_{\underbrace{\scriptstyle 1,\ldots,1}_{p-n} ,k_{p+1},\ldots}(N)
+\sum_{n=1}^{p-1}F^{n-1,2}_{\underbrace{\scriptstyle 1,\ldots,1}_{p-n-1},k_{p+1},\ldots}(N)\\
&+F^{p-1,{\rm sgn}(k_{p+1})(|k_{p+1}|+1)}_{k_{p+2},\ldots}(N)
+F^{p,k_{p+1}}_{k_{p+2},\ldots}(N),
\end{split}
\label{anconofs1pkpp1etc}
\ee
which can be analytically continued and 
calculated as a large $|N|$ expansion according to the method of section \ref{largeNk1ne1}.
This requires that the
$S_{\underbrace{\scriptstyle 1,\ldots,1}_n}(N)$ for $n=1,\ldots,p$ are known in the form in eq.\ (\ref{SeriesforSk2k3n}).
(These types of harmonic sums can also be determined by writing them in terms
of harmonic sums of the same type but of lower depth according to formulae presented in Ref.\ \cite{Blumlein:1998if}.
This gives polynomial representations for $S_{\underbrace{\scriptstyle 1,\ldots,1}_n}(N)$ as determinants.
However, for completeness we show how they can be determined within our approach.)
To find these for $n$ even, 
we use the result that follows from eq.\ (\ref{masterdecompofsp1}):
\be
S_{\underbrace{\scriptstyle 1,\ldots,1}_n}(N)=\frac{1}{2}\sum_{m=1}^{n-1} \left[(-1)^{m-1}
S_{\underbrace{\scriptstyle 1,\ldots,1}_m}(N) S_{\underbrace{\scriptstyle 1,\ldots,1}_{n-m}}(N)
+ F^{m-1,2}_{\underbrace{\scriptstyle 1,\ldots,1}_{n-m-1}}(N)\right] \ {\rm for}\ n\ {\rm even}.
\label{S1peven}
\ee
To obtain $S_{\underbrace{\scriptstyle 1,\ldots,1}_n}(N)$ with $n$ odd, we write it as
$\sum_{n_1=1}^N S_{\underbrace{\scriptstyle 1,\ldots,1}_{n-1}}(n_1)/n_1$,
where $S_{\underbrace{\scriptstyle 1,\ldots,1}_{n-1}}(n_1)$ is expanded as in eq.\ (\ref{S1peven}) with $N\rightarrow n_1$.
Equation (\ref{replacesum}) is used repetitively on the second term, which then becomes
\be
\begin{split}
\sum_{n_1=1}^N \frac{1}{n_1}S_{\underbrace{\scriptstyle 1,\ldots,1}_m}(n_1)S_{\underbrace{\scriptstyle 1,\ldots,1}_{n-m-1}}(n_1)
=&S_{\underbrace{\scriptstyle 1,\ldots,1}_{n-m}}(N)S_{\underbrace{\scriptstyle 1,\ldots,1}_m}(N)
+\sum_{n_1=1}^N \frac{1}{n_1^2}S_{\underbrace{\scriptstyle 1,\ldots,1}_{m-1}}(n_1)S_{\underbrace{\scriptstyle 1,\ldots,1}_{n-m-1}}(n_1)\\
&-\sum_{n_1=1}^N \frac{1}{n_1}S_{\underbrace{\scriptstyle 1,\ldots,1}_{m-1}}(n_1)S_{\underbrace{\scriptstyle 1,\ldots,1}_{n-m}}(n_1)\\
=&\sum_{k=1}^m (-1)^{k-1}\left(S_{\underbrace{\scriptstyle 1,\ldots,1}_{n-m+k-1}}(N)S_{\underbrace{\scriptstyle 1,\ldots,1}_{m-k+1}}(N)
+\sum_{n_1=1}^N \frac{1}{n_1^2}S_{\underbrace{\scriptstyle 1,\ldots,1}_{m-k}}(n_1)S_{\underbrace{\scriptstyle 1,\ldots,1}_{n-m+k-2}}(n_1)\right)\\
&+(-1)^m S_{\underbrace{\scriptstyle 1,\ldots,1}_n}(N),
\end{split}
\ee
and eq.\ (\ref{replacesum}) is used once on the first term in eq.\ (\ref{S1peven}).
The result is
\be
\begin{split}
S_{\underbrace{\scriptstyle 1,\ldots,1}_n}(N)=&\frac{1}{n}\sum_{m=1}^{n-2}
\Bigg[\sum_{k=1}^m \left((-1)^{m-k}S_{\underbrace{\scriptstyle 1,\ldots,1}_{n-m+k-1}}(N)S_{\underbrace{\scriptstyle 1,\ldots,1}_{m-k+1}}(N)
+F^{m-k,2}_{\underbrace{\scriptstyle 1,\ldots,1}_{n-m+k-2}}(N)
\right)\\
&+S_1(N) F^{m-1,2}_{\underbrace{\scriptstyle 1,\ldots,1}_{n-m-2}}(N)+F^{m-1,3}_{\underbrace{\scriptstyle 1,\ldots,1}_{n-m-2}}(N)
-(-1)^{m-1}\sum_{n_1=1}^N \frac{1}{n_1^2}S_{\underbrace{\scriptstyle 1,\ldots,1}_{m-1}}(n_1)S_{\underbrace{\scriptstyle 1,\ldots,1}_{n-m-2}}(n_1)S_1(n_1)\Bigg]\\
&{\rm for}\ n\geq 3\ {\rm and\ odd}.
\end{split}
\label{S1podd}
\ee

The quantity $F^{n,r}_{k_1,k_2,\ldots}(N)$ in eq.\ (\ref{defFnrks}) reduces to a harmonic sum
when $n=0$ and/or the indices $k_1,k_2,\ldots$ do not exist.
Consequently, provided $p$ is not too large, eq.\ (\ref{anconofs1pkpp1etc}) can be used to
easily express $S_{\underbrace{\scriptstyle 1,\ldots,1}_p ,k_{p+1},\ldots}(N)$
in terms of other harmonic sums as is done in Ref.\ \cite{Blumlein:1998if}.
For example, for $p=2$, omitting the argument $N$ for brevity,
\be
S_{1,1,k}=S_1 S_{1,k}-S_{1,1} S_k+S_{2,k}-S_{{\rm sgn}(k)(|k|+1),1}+S_{k,1,1}.
\ee
For large values of $p$, this expansion into other harmonic sums becomes more difficult. 
In any case, this is not the goal of this letter.

\section{Calculation when $|N|$ is not large \label{Nnotlarge}}

If $|N|$ is not sufficiently large for the evaluation of a harmonic sum as a large $|N|$ expansion to be sufficiently accurate,
the harmonic sum can be calculated in terms of harmonic sums whose arguments are large.
We begin with the relation between $S_{k_1,k_2,k_3,\ldots}(N)$ and $S_{k_1,k_2,k_3,\ldots}(N+r)$, 
where $r$ is any integer chosen such that $|N+r|$ is large, that follows from eq.\ (\ref{defofSk1k2forintN}):
\be
S_{k_1,k_2,k_3,\ldots}(N)=S_{k_1,k_2,k_3,\ldots}(N+r)-\sum_{n_1=1}^r\frac{({\rm sgn}(k_1))^{n_1+N}}{(n_1+N)^{|k_1|}}
S_{k_2,k_3,\ldots}(n_1+N),
\label{SNfromSNpr}
\ee
Applying eq.\ (\ref{SNfromSNpr}) to $S_{k_2,k_3,\ldots}(n_1+N)$, and then to $S_{k_3,\ldots}(n_1+N)$, and so on gives
\be
\begin{split}
S_{k_1,\ldots,k_p}(N)=S_{k_1,\ldots,k_p}(N+r)
-\sum_{n_1=1}^r &\frac{({\rm sgn}(k_1))^{n_1+N}}{(n_1+N)^{|k_1|}}\bigg[S_{k_2,\ldots,k_p}(N+r)\\
-\sum_{n_2=1}^{r-n_1} &\frac{({\rm sgn}(k_2))^{n_1+n_2+N}}{(n_1+n_2+N)^{|k_2|}}\bigg[S_{k_3,\ldots,k_p}(N+r)
-\ldots\\
-\sum_{n_p=1}^{r-n_1-\ldots-n_{p-1}} &\frac{({\rm sgn}(k_p))^{n_1+\ldots+n_p+N}}{(n_1+\ldots+n_p+N)^{|k_p|}}
\bigg]\ldots\bigg].
\end{split}
\label{SNfromSNpr2}
\ee
The harmonic sums $S_{k_i,k_{i+1},\ldots}(N+r)$ appearing in eq.\ (\ref{SNfromSNpr2}) 
can all be calculated according to the method of section \ref{largeNk1ne1} or \ref{largeNk1eq1},
i.e.\ as an expansion in $1/(N+r)$.

\section{Summary \label{summary}}

Our approach for evaluating a harmonic sum for any complex value of its argument
can be summarized as follows. We first consider the case of large $|N|$.
When $k_1 \neq1$, $S_{k_1,k_2,k_3,\ldots}(N)$ in the form in eq.\ (\ref{SeriesforSk2k3n}) can be obtained
from $S_{k_2,k_3,\ldots}(N)$ in the form in eq.\ (\ref{SeriesforSk2k3n}) by analytically
continuing $S_{k_1,k_2,k_3,\ldots}(N)$ to complex $N$ using eq.\ (\ref{defofSk1k2forcompN}),
writing $S_{k_2,k_3,\ldots}(n_1+N)$ in the form in eq.\ (\ref{SeriesforSk2k3n}) with $N\rightarrow n_1+N$, 
making the identifications in eq.\ (\ref{genzetaident}), and finally using the expansion in eq.\ (\ref{expanzeta})
to put $S_{k_1,k_2,\ldots}(N)$ in the form in eq.\ (\ref{SeriesforSk2k3n}).
Since the $S_k(N)$ are known in the form in eqs.\ (\ref{Snintermsofpsiderivs}), (\ref{Skcont}) and (\ref{Snintermsofpsiderivs2}),
where $\psi(N+1)$ has the large $|N|$ expansion in eq.\ (\ref{psiatlargeN}), this means that all
harmonic sums for which $k_1 \neq1$ can be put in the form in eq.\ (\ref{SeriesforSk2k3n}).
Harmonic sums for which $k_1 =1$ can be put in the form in eq.\ (\ref{SeriesforSk2k3n}) in a similar manner.
$S_{\underbrace{\scriptstyle 1,\ldots,1}_p ,k_{p+1},\ldots}(N)$ can be written in the form in eq.\ (\ref{anconofs1pkpp1etc}),
where $F^{n,r}_{k_1,k_2,\ldots}(N)$ can be analytically continued and expanded in $1/N$ 
in precisely the same way as $S_{k_1,k_2,k_3,\ldots}(N)$ for $k_1 \neq 1$ was,
and can therefore be written in the form in eq.\ (\ref{SeriesforSk2k3n}).
This requires that the
$S_{\underbrace{\scriptstyle 1,\ldots,1}_n}(N)$ for $n=1,\ldots,p$ are known in the form in eq.\ (\ref{SeriesforSk2k3n}),
which are found using eqs.\ (\ref{S1peven}) and (\ref{S1podd}).
When $|N|$ is not large, $S_{k_1,k_2,k_3,\ldots}(N)$ can be calculated from harmonic sums with arbitrarily large argument
using eq.\ (\ref{SNfromSNpr2}).

For convenience, FORTRAN and Mathematica files for calculating all harmonic sums of 
weight $\leq 5$ to double precision accuracy are provided at
\verb+http://www.desy.de/~simon/HarmonicSums+, which are suitable for leading order
(LO), next-to-leading order (NLO) and next-to-next-to-leading order (NNLO) calculations in Mellin space.
The extension of these programs to higher weights and higher accuracies is obvious.

A preprint recently appeared \cite{Blumlein:2009ta} 
which derives relations between harmonic sums of different weight.
This allows harmonic sums up to a given weight to be calculated in terms of a few basic functions,
each of which can be represented as a large $|N|$ expansion.

\begin{acknowledgments}

The author thanks Alex Mitov and Oleg Veretin for useful discussions.

\end{acknowledgments}

\newpage
\appendix

\section{Explicit results for large $|N|$}

In this appendix we present explicit results for each harmonic sum with weight less than or equal to
3 as series in the inverse of its argument up to $O(1/N^{10})$ determined using the method in this letter.
We define $\tilde{\ln}(N)=\ln (N)+\gamma_E$ and $\eta(N)=(-1)^N$.
The constants $c_{1,\pm 1,\pm 1}$ appearing in $S_{1,\pm 1,\pm1}(N)$ can be evaulated numerically as
described in the last paragraph of section \ref{largeNk1ne1}.
For example, they are given to 5 decimal places by $c_{1,-1,1}=-0.30883$, $c_{1,1,-1}=0.51459$ and $c_{1,-1,-1}=-0.37070$.
The constants $S_{-1,\pm 1,\pm 1}(\infty)$ may also be evaluated numerically this way, and one finds $S_{-1,-1,1}(\infty)=1.47800$,
$S_{-1,1,-1}(\infty)=0.66484$ and $S_{-1,1,1}(\infty)=-0.53721$.
Both the constants $c_{1,\pm 1,\pm 1}$ and $S_{-1,\pm 1,\pm 1}(\infty)$ may also be expressed in terms of rational and 
special numbers by using the results obtained in Ref.\ \cite{Blumlein:1998if}.

\subsection{Weight 1}

\be
S_{-1}(N)=\eta \left(\frac{1}{2 N}-\frac{1}{4 N^2}+\frac{1}{8 N^4}-\frac{1}{4 N^6}+\frac{17}{16 N^8}-\frac{31}{4 N^{10}}\right)-\ln 2 
\ee
\be
S_1(N)=\tilde{\ln} N +\frac{1}{2 N}-\frac{1}{12 N^2}+\frac{1}{120 N^4}-\frac{1}{252 N^6}+\frac{1}{240 N^8}-\frac{1}{132 N^{10}}
\ee

\subsection{Weight 2}

\be
S_{-2}(N)=\eta \left(\frac{1}{2 N^2}-\frac{1}{2 N^3}+\frac{1}{2 N^5}-\frac{3}{2 N^7}+\frac{17}{2 N^9}\right)-\frac{\pi ^2}{12}
\ee
\be
S_2(N)=\frac{\pi ^2}{6}-\frac{1}{N}+\frac{1}{2 N^2}-\frac{1}{6 N^3}+\frac{1}{30 N^5}-\frac{1}{42 N^7}+\frac{1}{30 N^9}
\ee
\be
\begin{split}
S_{-1,-1}(N)=&\eta \ln 2  \left(-\frac{1}{2 N}+\frac{1}{4 N^2}-\frac{1}{8 N^4}+\frac{1}{4 N^6}-\frac{17}{16 N^8}+\frac{31}{4 N^{10}}\right)+\frac{\ln^2 2 }{2}+\frac{\pi ^2}{12}\\
&-\frac{1}{2 N}+\frac{3}{8 N^2}-\frac{5}{24 N^3}+\frac{1}{32 N^4}+\frac{19}{240 N^5}-\frac{1}{32 N^6}-\frac{23}{168 N^7}+\frac{9}{128 N^8}+\frac{263}{480 N^9}-\frac{19}{64 N^{10}}
\end{split}
\ee
\be
\begin{split}
S_{-1,1}(N)=&\eta \left(\frac{1}{2 N^2}-\frac{7}{24 N^3}-\frac{1}{6 N^4}+\frac{61}{240 N^5}+\frac{73}{160 N^6}-\frac{379}{504 N^7}-\frac{11461}{5040 N^8}+\frac{2041}{480 N^9}+\frac{371417}{20160 N^{10}}\right)\\
&+\eta \left(\frac{1}{2 N}-\frac{1}{4 N^2}+\frac{1}{8 N^4}-\frac{1}{4 N^6}+\frac{17}{16 N^8}-\frac{31}{4 N^{10}}\right) \tilde{\ln} N +\frac{\ln^2 2 }{2}-\frac{\pi ^2}{12}
\end{split}
\ee
\be
\begin{split}
S_{1,-1}(N)=&\eta \left(\frac{1}{4 N^2}-\frac{3}{8 N^3}+\frac{3}{16 N^4}+\frac{5}{16 N^5}-\frac{15}{32 N^6}-\frac{7}{8 N^7}+\frac{147}{64 N^8}+\frac{153}{32 N^9}-\frac{1185}{64 N^{10}}\right)\\
&+\ln 2  \left(-\frac{1}{2 N}+\frac{1}{12 N^2}-\frac{1}{120 N^4}+\frac{1}{252 N^6}-\frac{1}{240 N^8}+\frac{1}{132 N^{10}}\right)-\ln 2  \tilde{\ln} N-\frac{\ln^2 2 }{2}
\end{split}
\ee
\be
\begin{split}
S_{1,1}(N)=&\frac{\tilde{\ln}^2 N }{2}+\left(\frac{1}{2 N}-\frac{1}{12 N^2}+\frac{1}{120 N^4}-\frac{1}{252 N^6}+\frac{1}{240 N^8}-\frac{1}{132 N^{10}}\right) \tilde{\ln} N +\frac{\pi ^2}{12}\\
&-\frac{1}{2 N}+\frac{3}{8 N^2}-\frac{1}{8 N^3}+\frac{1}{288 N^4}+\frac{1}{48 N^5}-\frac{1}{1440 N^6}-\frac{1}{72 N^7}+\frac{221}{604800 N^8}+\frac{3}{160 N^9}-\frac{23}{60480 N^{10}}
\end{split}
\ee

\newpage
\subsection{Weight 3}

\be
S_{-3}(N)=\eta \left(\frac{1}{2N^3}-\frac{3}{4N^4}+\frac{5}{4N^6}-\frac{21}{4N^8}+\frac{153}{4N^{10}}\right)-\frac{3\zeta (3)}{4} 
\ee
\be
S_3(N)=\zeta (3)-\frac{1}{2 N^2}+\frac{1}{2 N^3}-\frac{1}{4 N^4}+\frac{1}{12 N^6}-\frac{1}{12 N^8}+\frac{3}{20 N^{10}}
\ee
\be
\begin{split}
S_{-2,-1}(N)=&\eta \left(-\frac{1}{2 N^2}+\frac{1}{2 N^3}-\frac{1}{2 N^5}+\frac{3}{2 N^7}-\frac{17}{2 N^9}\right) \ln 2 +\frac{\pi ^2 \ln 2 }{4}-\frac{5 \zeta (3)}{8}\\
&-\frac{1}{4 N^2}+\frac{1}{3 N^3}-\frac{1}{4 N^4}+\frac{7}{120 N^5}+\frac{5}{48 N^6}-\frac{23}{336 N^7}-\frac{1}{6 N^8}+\frac{13}{80 N^9}+\frac{97}{160 N^{10}}
\end{split}
\ee
\be
\begin{split}
S_{-2,1}(N)=&\eta \left(\frac{1}{2 N^3}-\frac{5}{12 N^4}-\frac{11}{24 N^5}+\frac{151}{240 N^6}+\frac{469}{240 N^7}-\frac{331}{126 N^8}-\frac{67379}{5040 N^9}+\frac{9181}{480 N^{10}}\right)\\
&+\eta \left(\frac{1}{2 N^2}-\frac{1}{2 N^3}+\frac{1}{2 N^5}-\frac{3}{2 N^7}+\frac{17}{2 N^9}\right) \tilde{\ln} N -\frac{5 \zeta (3)}{8}
\end{split}
\ee
\be
\begin{split}
S_{2,-1}(N)=&\eta \left(\frac{1}{4 N^3}-\frac{1}{2 N^4}+\frac{1}{4 N^5}+\frac{11}{16 N^6}-\frac{13}{16 N^7}-\frac{11}{4 N^8}+\frac{39}{8 N^9}+\frac{629}{32 N^{10}}\right)\\
&+\ln 2  \left(\frac{1}{N}-\frac{1}{2 N^2}+\frac{1}{6 N^3}-\frac{1}{30 N^5}+\frac{1}{42 N^7}-\frac{1}{30 N^9}\right)-\frac{\pi ^2}{4}\ln 2+\frac{\zeta (3)}{4}
\end{split}
\ee
\be
\begin{split}
S_{2,1}(N)=&\left(-\frac{1}{N}+\frac{1}{2 N^2}-\frac{1}{6 N^3}+\frac{1}{30 N^5}-\frac{1}{42 N^7}+\frac{1}{30 N^9}\right) \tilde{\ln} N+2 \zeta (3) \\
&-\frac{1}{N}-\frac{1}{4 N^2}+\frac{13}{36 N^3}-\frac{1}{6 N^4}-\frac{1}{100 N^5}+\frac{11}{240 N^6}+\frac{601}{35280 N^7}-\frac{11}{252 N^8}-\frac{247}{7560 N^9}+\frac{37}{480 N^{10}}
\end{split}
\ee
\be
\begin{split}
S_{-1,-2}(N)=&-\frac{\pi ^2 \ln 2 }{6}+\frac{13 \zeta (3)}{8}+\eta \left(-\frac{\pi ^2}{24 N}+\frac{\pi ^2}{48 N^2}-\frac{\pi ^2}{96 N^4}+\frac{\pi ^2}{48 N^6}-\frac{17 \pi ^2}{192 N^8}+\frac{31 \pi ^2}{48 N^{10}}\right)\\
&-\frac{1}{4 N^2}+\frac{5}{12 N^3}-\frac{3}{8 N^4}+\frac{1}{15 N^5}+\frac{7}{24 N^6}-\frac{5}{42 N^7}-\frac{19}{24 N^8}+\frac{2}{5 N^9}+\frac{173}{40 N^{10}}
\end{split}
\ee
\be
\begin{split}
S_{-1,2}(N)=&\eta \bigg(\frac{\pi ^2}{12 N}-\frac{12+\pi^2}{24 N^2}+\frac{3}{4 N^3}
-\frac{22-\pi ^2}{48 N^4}-\frac{1}{3 N^5}+\frac{77-5\pi ^2}{120 N^6}\\
&+\frac{31}{30 N^7}
-\frac{1772-119 \pi ^2}{672 N^8}-\frac{206}{35 N^9}+\frac{2297-155 \pi ^2}{120 N^{10}}\bigg)
+\frac{\pi ^2 \ln 2 }{12}-\zeta (3)
\end{split}
\ee
\be
\begin{split}
S_{1,-2}(N)=&\eta \left(\frac{1}{4 N^3}-\frac{5}{8 N^4}+\frac{1}{2 N^5}+\frac{7}{8 N^6}-\frac{2}{N^7}-\frac{27}{8 N^8}+\frac{27}{2 N^9}+\frac{187}{8 N^{10}}\right)-\frac{\pi ^2 \tilde{\ln} N }{12}-\frac{\zeta (3)}{8}\\
&+\pi ^2\left(-\frac{1}{24 N}+\frac{1}{144 N^2}-\frac{1}{1440 N^4}+\frac{1}{3024 N^6}-\frac{1}{2880 N^8}+\frac{1}{1584 N^{10}}\right)
\end{split}
\ee
\be
\begin{split}
S_{1,2}(N)=&\frac{\pi ^2 \tilde{\ln} N }{6}-\zeta (3)
+\frac{1+\frac{\pi ^2}{12}}{N}+\frac{-\frac{3}{4}-\frac{\pi ^2}{72}}{N^2}+\frac{17}{36 N^3}+\frac{-150+\pi ^2}{720 N^4}+\frac{7}{450 N^5}\\
&+\frac{\frac{7}{120}-\frac{\pi ^2}{1512}}{N^6}-\frac{38}{2205 N^7}+\frac{-\frac{3}{56}+\frac{\pi ^2}{1440}}{N^8}+\frac{11}{350 N^9}+\frac{\frac{11}{120}-\frac{\pi ^2}{792}}{N^{10}}
\end{split}
\ee
\be
\begin{split}
S_{-1,-1,-1}(N)=&+\ln 2 \left(\frac{1}{2 N}-\frac{3}{8 N^2}+\frac{5}{24 N^3}-\frac{1 }{32 N^4}-\frac{19}{240 N^5}+\frac{1}{32 N^6}+\frac{23}{168 N^7}-\frac{9}{128 N^8}-\frac{263}{480 N^9}+\frac{19}{64 N^{10}}\right)\\
&+\eta  \bigg[\ln^2 2 \left(\frac{1}{4 N}-\frac{1}{8 N^2}+\frac{1}{16 N^4}-\frac{1}{8 N^6}+\frac{17}{32 N^8}-\frac{31}{8 N^{10}}\right)
\\
&-\frac{1}{4 N^2}+\frac{7}{16 N^3}-\frac{37}{96 N^4}-\frac{5}{192 N^5}+\frac{901}{1920 N^6}+\frac{91}{960 N^7}-\frac{9923}{5376 N^8}-\frac{4213}{8960 N^9}+\frac{100969}{7680 N^{10}}\\
&+\pi ^2\left(\frac{1}{24 N}-\frac{1}{48 N^2}+\frac{1}{96 N^4}-\frac{1}{48 N^6}+\frac{17}{192 N^8}-\frac{31}{48 N^{10}}\right)\bigg]
-\frac{\ln ^3 2 }{6}-\frac{\zeta (3)}{4}
\end{split}
\ee
\be
\begin{split}
S_{-1,-1,1}(N)=&\tilde{\ln} N \left(-\frac{1}{2 N}+\frac{3}{8 N^2}-\frac{5}{24 N^3}+\frac{1}{32 N^4}+\frac{19}{240 N^5}-\frac{1}{32 N^6}-\frac{23}{168 N^7}+\frac{9}{128 N^8}+\frac{263}{480 N^9}-\frac{19}{64 N^{10}}\right)\\
&+\eta  \bigg[\ln^2 2 \left(\frac{1}{4 N}-\frac{1}{8 N^2}+\frac{1}{16 N^4}-\frac{1}{8 N^6}+\frac{17}{32 N^8}-\frac{31}{8 N^{10}}\right)\\
&+\pi ^2\left(-\frac{1}{24 N}+\frac{1}{48 N^2}-\frac{1}{96 N^4}+\frac{1}{48 N^6}-\frac{17}{192 N^8}+\frac{31}{48 N^{10}}\right)\bigg]\\
&-\frac{1}{2 N}-\frac{3}{16 N^2}+\frac{7}{18 N^3}-\frac{33}{128 N^4}-\frac{11}{200 N^5}+\frac{47}{240 N^6}+\frac{4999}{28224 N^7}\\
&-\frac{56657}{107520 N^8}-\frac{28829}{30240 N^9}+\frac{112723}{40320 N^{10}}+S_{-1,-1,1}(\infty)
\end{split}
\ee
\be
\begin{split}
S_{-1,1,-1}(N)=&\eta  \bigg[\tilde{\ln} N  \ln 2 \left(-\frac{1}{2 N}+\frac{1}{4 N^2}-\frac{1}{8 N^4}+\frac{1}{4 N^6}-\frac{17}{16 N^8}+\frac{31}{4 N^{10}}\right)\\
&+\ln 2 \left(-\frac{1}{2 N^2}+\frac{7}{24 N^3}+\frac{1}{6 N^4}-\frac{61}{240 N^5}-\frac{73}{160 N^6}+\frac{379}{504 N^7}+\frac{11461}{5040 N^8}-\frac{2041}{480 N^9}-\frac{371417 }{20160 N^{10}}\right)\\
&+\ln^2 2 \left(-\frac{1}{4 N}+\frac{1}{8 N^2}-\frac{1}{16 N^4}+\frac{1}{8 N^6}-\frac{17}{32 N^8}+\frac{31}{8 N^{10}}\right)\bigg]\\
&-\frac{1}{8 N^2}+\frac{1}{4 N^3}-\frac{19}{64 N^4}+\frac{5}{32 N^5}+\frac{17}{96 N^6}-\frac{21}{64 N^7}-\frac{641}{1536 N^8}+\frac{183}{128 N^9}+\frac{2733}{1280 N^{10}}+S_{-1,1,-1}(\infty)
\end{split}
\ee
\be
\begin{split}
S_{-1,1,1}(N)=&\eta  \bigg[\tilde{\ln}^2 N \left(\frac{1}{4 N}-\frac{1}{8 N^2}+\frac{1}{16 N^4}-\frac{1}{8 N^6}+\frac{17 }{32 N^8}-\frac{31 }{8 N^{10}}\right)\\
&+\tilde{\ln} N \left(\frac{1}{2 N^2}-\frac{7 }{24 N^3}-\frac{1}{6 N^4}+\frac{61 }{240 N^5}+\frac{73 }{160 N^6}-\frac{379 }{504 N^7}-\frac{11461}{5040 N^8}+\frac{2041}{480 N^9}+\frac{371417 }{20160 N^{10}}\right)\\
&-\frac{1}{4 N^2}+\frac{9}{16 N^3}-\frac{23}{96 N^4}-\frac{227}{576 N^5}+\frac{517}{5760 N^6}+\frac{4301}{2880 N^7}+\frac{8371}{26880 N^8}-\frac{11644939}{1209600 N^9}-\frac{15388481}{2419200 N^{10}}\\
&+\pi ^2\left(\frac{1}{24 N}-\frac{1}{48 N^2}+\frac{1}{96 N^4}-\frac{1}{48 N^6}+\frac{17}{192 N^8}-\frac{31}{48 N^{10}}\right)\bigg]+S_{-1,1,1}(\infty)
\end{split}
\ee
\be
\begin{split}
S_{1,-1,-1}(N)=&\eta \ln 2 \left(-\frac{1 }{4 N^2}+\frac{3  }{8 N^3}-\frac{3 }{16 N^4}-\frac{5 }{16 N^5}+\frac{15 }{32 N^6}+\frac{7 }{8 N^7}-\frac{147 }{64 N^8}-\frac{153 }{32 N^9}+\frac{1185 }{64 N^{10}}\right)\\
&+\tilde{\ln} N \left(\frac{\pi ^2 }{12}+\frac{ \ln^2 2 }{2}\right)+\ln^2 2 \left(\frac{1}{4 N}-\frac{1 }{24 N^2}+\frac{1 }{240 N^4}-\frac{1 }{504 N^6}+\frac{1 }{480 N^8}-\frac{1 }{264 N^{10}}\right)\\
&+\pi ^2\left(\frac{1}{24 N}-\frac{1}{144 N^2}+\frac{1}{1440 N^4}-\frac{1}{3024 N^6}+\frac{1}{2880 N^8}-\frac{1}{1584 N^{10}}\right)\\
&+\frac{1}{2 N}-\frac{7}{16 N^2}+\frac{49}{144 N^3}-\frac{79}{384 N^4}+\frac{757}{14400 N^5}\\
&+\frac{121}{1920 N^6}-\frac{8251}{141120 N^7}-\frac{1745}{21504 N^8}+\frac{5919}{44800 N^9}+\frac{2071}{7680 N^{10}}+c_{1,-1,-1}
\end{split}
\ee
\be
\begin{split}
S_{1,-1,1}(N)=&  \tilde{\ln} N \left(\frac{\ln^2 2}{2}-\frac{\pi ^2}{12}\right)\\
&+\eta  \bigg[\tilde{\ln} N \left(\frac{1}{4 N^2}-\frac{3 }{8 N^3}+\frac{3 }{16 N^4}+\frac{5}{16 N^5}-\frac{15}{32 N^6}-\frac{7}{8 N^7}+\frac{147}{64 N^8}+\frac{153}{32 N^9}-\frac{1185 }{64 N^{10}}\right)\\
&+\frac{3}{8 N^3}-\frac{7}{12 N^4}-\frac{1}{16 N^5}+\frac{1187}{960 N^6}+\frac{197}{960 N^7}-\frac{127091}{20160 N^8}-\frac{15}{14 N^9}+\frac{179131}{3360 N^{10}}\bigg]\\
&+\ln^2 2 \left(\frac{1}{4 N}-\frac{1}{24 N^2}+\frac{1}{240 N^4}-\frac{1}{504 N^6}+\frac{1}{480 N^8}-\frac{1}{264 N^{10}}\right)\\
&+\pi ^2\left(-\frac{1}{24 N}+\frac{1}{144 N^2}-\frac{1}{1440 N^4}+\frac{1}{3024 N^6}-\frac{1}{2880 N^8}+\frac{1}{1584 N^{10}}\right)+c_{1,-1,1}
\end{split}
\ee
\be
\begin{split}
S_{1,1,-1}(N)=&-\frac{\tilde{\ln}^2 N  \ln 2 }{2}-\frac{\tilde{\ln} N  \ln^2 2 }{2}+\tilde{\ln} N  \ln 2 \left(-\frac{1}{2 N}+\frac{1}{12 N^2}-\frac{1}{120 N^4}+\frac{1}{252 N^6}-\frac{1}{240 N^8}+\frac{1}{132 N^{10}}\right)\\
&+\ln 2 \bigg(\frac{1}{2 N}-\frac{3}{8 N^2}+\frac{1}{8 N^3}-\frac{1}{288 N^4}-\frac{1}{48 N^5}\\
&+\frac{1}{1440 N^6}+\frac{1}{72 N^7}-\frac{221}{604800 N^8}-\frac{3 }{160 N^9}+\frac{23}{60480 N^{10}}\bigg)\\
&-\frac{\pi ^2 \ln 2 }{6}+\eta  \left(\frac{1}{8 N^3}-\frac{3}{8 N^4}+\frac{15}{32 N^5}+\frac{15}{64 N^6}-\frac{105}{64 N^7}-\frac{7}{64 N^8}+\frac{1323}{128 N^9}-\frac{1035}{256 N^{10}}\right)\\
&+\ln^2 2 \left(-\frac{1}{4 N}+\frac{1}{24 N^2}-\frac{1}{240 N^4}+\frac{1}{504 N^6}-\frac{1}{480 N^8}+\frac{1}{264 N^{10}}\right)+c_{1,1,-1}
\end{split}
\ee
\be
\begin{split}
S_{1,1,1}(N)=&\frac{\tilde{\ln} ^3 N}{6}+\frac{\pi ^2 \tilde{\ln} N }{12}+\tilde{\ln}^2 N \left(\frac{1}{4 N}-\frac{1}{24 N^2}+\frac{1}{240 N^4}-\frac{1}{504 N^6}+\frac{1}{480 N^8}-\frac{1}{264 N^{10}}\right)\\
&+\tilde{\ln} N \bigg(-\frac{1}{2 N}+\frac{3}{8 N^2}-\frac{1}{8 N^3}+\frac{1}{288 N^4}+\frac{1}{48 N^5}\\
&-\frac{1}{1440 N^6}-\frac{1}{72 N^7}+\frac{221}{604800 N^8}+\frac{3 }{160 N^9}-\frac{23}{60480 N^{10}}\bigg)\\
&-\frac{5}{12 N^2}+\frac{17}{48 N^3}-\frac{5}{32 N^4}+\frac{13}{2880 N^5}+\frac{2029}{51840 N^6}\\
&-\frac{1}{2240 N^7}-\frac{8513}{241920 N^8}-\frac{59}{134400 N^9}+\frac{434579}{7257600 N^{10}}\\
&+\pi ^2\left(\frac{1}{24 N}-\frac{1}{144 N^2}+\frac{1}{1440 N^4}-\frac{1}{3024 N^6}+\frac{1}{2880 N^8}-\frac{1}{1584 N^{10}}\right)+\frac{\zeta (3)}{3}
\end{split}
\ee

\end{document}